\documentclass[10pt,journal,letterpaper]{IEEEtran}
\usepackage{latexsym}
\usepackage{graphics}
\usepackage{epstopdf}
\usepackage{epsfig}
\usepackage{amsfonts}
\usepackage{float}
\usepackage{amssymb}
\usepackage{stfloats}
\usepackage{multirow}
\usepackage{multicol}
\usepackage{amsmath}
\usepackage{bm}
\usepackage{array}
\usepackage{cite}
\usepackage{tabularx}
\usepackage{enumerate}
\usepackage{times}
\usepackage[noend]{algpseudocode}
\usepackage{threeparttable}
\usepackage{amsfonts, amssymb, amsthm}
\usepackage{latexsym, graphicx}
\usepackage{epsfig}
\usepackage{float}
\usepackage{fancyhdr}
\usepackage{subfigure}
\usepackage{graphicx}
\usepackage{color}
\usepackage{amssymb}
\usepackage{amsfonts}
\usepackage{subfigure}
\usepackage{balance}
\usepackage{tabularx}
\usepackage{bm}
\usepackage{multirow}
\usepackage{graphicx,times,amsmath}
\usepackage{amsfonts}
\usepackage{mathrsfs}
\usepackage[table]{xcolor}
\usepackage[noend]{algpseudocode}
\usepackage{algorithmicx,algorithm}

\begin{document}
\vskip -10ex
\setlength{\abovecaptionskip}{0pt}
\setlength{\belowcaptionskip}{0pt}
\intextsep=5pt
\textfloatsep=5pt
\title{Branching Dueling Q-Network Based Online Scheduling of a Microgrid With Distributed Energy Storage Systems}
\author{Hang Shuai,~\IEEEmembership{Member,~IEEE},~Fangxing (Fran) Li,~\IEEEmembership{Fellow,~IEEE},~H\'{e}ctor Pulgar-Painemal,~\IEEEmembership{Senior Member,~IEEE},~and Yaosuo Xue,~\IEEEmembership{Senior Member,~IEEE}
\vspace{0.2cm}

\thanks{H. Shuai, F. Li, and H. Pulgar-Painemal are with the Department of Electrical Engineering and Computer Science, University of Tennessee, Knoxville, TN 37996 USA. \textit{Corresponding author: Fangxing (Fran) Li.}

Y. Xue is with Oak Ridge National Laboratory (ORNL), Oak Ridge, Tennessee, USA.
}
}

\IEEEaftertitletext{\vspace{-2.99\baselineskip}}
\markboth{IEEE TRANSACTIONS ON ,~Vol.~,
No.~,~~2021}{Shell \MakeLowercase{\textit{et al.}}: Bare Demo of
IEEEtran.cls for Journals}

\markboth{SUBMITTED TO IEEE JOURNAL FOR POSSIBLE PUBLICATION. COPYRIGHT WILL BE TRANSFERRED WITHOUT NOTICE}%
{Shell \MakeLowercase{\textit{et al.}}: Bare Demo of IEEEtran.cls for IEEE Transactions on Magnetics Journals}

\maketitle

\begin{abstract}
This letter investigates a Branching Dueling Q-Network (BDQ) based online operation strategy for a microgrid with distributed battery energy storage systems (BESSs) operating under uncertainties.
The developed deep reinforcement learning (DRL) based microgrid online optimization strategy can achieve a linear increase in the number of neural network outputs with the number of distributed BESSs, which overcomes the curse of dimensionality caused by the charge and discharge decisions of multiple BESSs.
Numerical simulations validate the effectiveness of the proposed method.
\end{abstract}

\vspace{-0.25cm}

\section{Introduction}
\PARstart{M}{icrogrid} is a promising concept for addressing the challenges of integrating distributed renewable energy and energy storage systems into power networks.
Online optimization, which schedules the operation of microgrids according to the real-time state of the system, is a key technique to ensure the economic operation of microgrids.

However, the uncertainties of renewable energy bring great challenges to the online optimization of microgrids.
To address this problem, researchers have proposed several online optimization methods, such as model predictive control (MPC) \cite{Gu}, and approximate dynamic programming (ADP) based algorithm \cite{Hang2019ADP}. 
Nevertheless, the online optimization performance of the above methods relies on forecasting information. 
So, the performance is affected by the forecasting accuracy of renewable energy and load power.
To decrease the dependence on forecasting, several other online optimization approaches for microgrids have been proposed, including the Lyapunov optimization \cite{Shi2017}, the CHASE algorithm \cite{jia2020novel}, and the recently developed deep reinforcement learning (DRL) based optimization methods (e.g., deep Q Network (DQN) \cite{franccois2016deep}, MuZero \cite{Hang2020MCTS}).

Compared with the conventional microgrid online optimization approaches (e.g., MPC), DRL based algorithms learn to operate the system via historical renewable power generation and load sequences, and can make near-optimal scheduling without using any forecasting information \cite{Hang2020MCTS}.
However, the above-mentioned works mainly focus on the online optimization of a microgrid with a single battery energy storage system (BESS), which fails to address the distributed location characteristic of BESSs.
With the rapid development of commercial and home energy storage techniques, plenty of BESSs will be installed in distributed locations of microgrids.
The huge action space introduced by multiple BESSs brings great challenges to the discrete-action based DRL optimization methods.
For instance, the number of actions that need to be explicitly represented in the DQN \cite{franccois2016deep} or Muzero \cite{Hang2020MCTS} based agents grows exponentially with an increasing number of BESSs.
As a result, the DRL based optimization approaches proposed in \cite{franccois2016deep, Hang2020MCTS} are difficult to adapt for a microgrid with distributed BESSs.

This letter develops a novel Branching Dueling Q-Network (BDQ) \cite{tavakoli2018action} based online optimization strategy for a microgrid with distributed BESSs.
The designed BDQ based intelligent agent contains a shared decision module followed by several network branches, one for each BESS.
The developed algorithm can achieve a linear increase of the number of neural network outputs with the number of distributed BESSs, which will provide great scalability and increase the applicability of the algorithm.
Specifically, to accommodate the characteristics of historical renewable energy power generation and load power sequences, a long short-term memory (LSTM) based shared decision module architecture is designed in this letter to extract features from historical data.

This letter is organized as follows.
Section II formulates the microgrid online optimization problem as a mixed integer second-order cone programming (MISOCP) problem by adopting a branch power flow model.
In Section III, the BDQ based online optimization algorithm for the microgrid is designed.
The numerical simulations are presented in Section IV.
Section V concludes this work.
\vspace{-0.20cm}

\section{Optimization Model of the Microgrid}
The microgrid investigated in this letter works in a grid connection mode and consists of electric loads, BESSs, controllable distributed generators (DGs) (e.g., diesel generators), and uncontrollable DGs (e.g., PV panel systems and wind turbines).
The goal of online optimization is to minimize the operation cost of the microgrid over the optimization horizon under the necessary constraints.
The objective function consists of the fuel cost of controllable DGs, the power exchange cost of the microgrid and the utility grid, the degradation cost of BESSs, and the renewable energy curtailment cost.
The operational constraints considered in this work include the power generation limit and ramp rate constraints, the power exchange limit between the microgrid and the utility, the charge/discharge power limit of BESSs, the branch power flow constraints, etc.
The details of the microgrid optimization model can be found in equation (1) - (25) of reference \cite{Hang2020MCTS}.

\section{BDQ Based Online Optimization Strategy for a Microgrid with Distributed BESSs}
\begin{figure}[h]\centering
\includegraphics[width=3.5in]{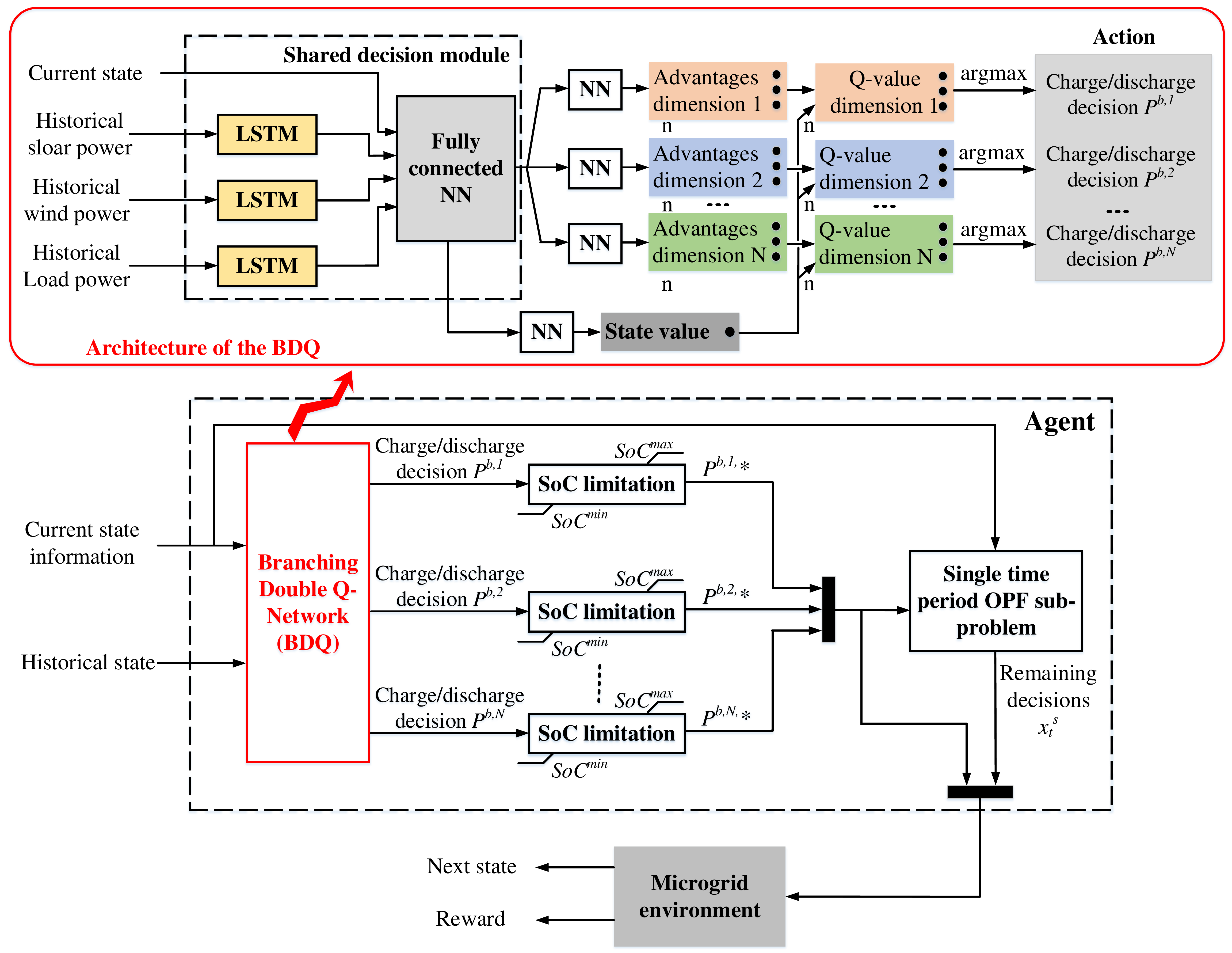}
\caption{The developed BDQ based online optimization strategy for a microgrid with distributed BESSs.} \label{fig:BDQ_Arch}
\vspace{-0.5em}
\end{figure}

The decision variables of the online optimization problem include the complex power generation of controllable DGs, PV panels, and wind turbines; the charge and discharge power of distributed BESSs; the complex power exchange between the microgrid and the utility grid; the branch current; the bus voltage; etc.
However, the high-dimensional continuous actions force us to face the curse of dimensionality when
applying the reinforcement learning methods to solving our problem.
To this end, we develop the BDQ \cite{tavakoli2018action} based online optimization approach for a microgrid with distributed BESSs, as illustrated in Fig. \ref{fig:BDQ_Arch}.
The BDQ agent only determines the charge and discharge power of distributed BESSs, while the remaining decisions are obtained by solving the single-time period optimal power flow (OPF) subproblem.
The advantage of the proposed optimization architecture is that it can operate the system without dependence on any renewable and load power prediction information.

The designed network architecture of the BDQ agent is also given in Fig. \ref{fig:BDQ_Arch}.
The shared decision module consists of three LSTM units and a fully connected network.
The LSTM units extract features from load power and renewable energy power sequences, then the extracted features concatenate with the current state of the microgrid and are then fed into a multilayer network.
The features computed by the shared decision module are then used to compute the state value and the state-dependent action advantages on the subsequent independent branches \cite{tavakoli2018action}. 
Note that each branch corresponds to a BESS in this work.
The state value and the state-dependent action advantages are combined and input to neural networks to compute the Q-values for each BESS charge and discharge dimension.
We discretize the charge and discharge decision of each BESS into $n$ feasible values.
The individual branch’s Q-value at state $s$ when taking decision $P_d^b$ can be given by:
\begin{equation}\label{EQ1}
Q_d(s, P_d^b) = V(s) + \big(A_d(s, P_d^b) - \frac{1}{n} \sum_{P_{d}^{b'} \in {\mathcal{X}_d}} A_d(s, P_{d}^{b'})\big)  
\end{equation} 
where, $d \in \{1, 2, \cdots, N\}$ represents the $d$th BESS; $V(s)$ is the state value output by the shared decision module; $A_d(s, P_d^b)$ is the state-dependent action advantage, and $P_d^b \in \mathcal{X}_d$.
$\mathcal{X}_d$ represents the feasible action space of the $d$th BESS.

The neural network weights of the BDQ agent are updated by minimizing the following loss function:
\begin{equation}\label{EQ2}
L =  \Xi_{(s, P^b, r,s^{'}) \sim D} \bigg[\frac{1}{N} \sum_d (y_d - Q_d (s, P_d^b))^2 \bigg]
\end{equation}
where, $y_d$ is the temporal-difference (TD) target for the BDQ agent, which can be computed by:
\begin{equation}\label{EQ3}
y_d = r + \gamma \frac{1}{N} \sum_d Q_d^{-} \big(s^{'}, \arg \max_{P_{d}^{b'} \in {\mathcal{X}_d}} Q_d(s^{'}, P_{d}^{b'})\big)
\end{equation} 
where, $r$ represents the reward after taking decision $P^b$; $\gamma$ is the discount factor.
The details of the training process of the developed BDQ based online optimization algorithm for a microgrid with multiple BESSs is shown in \textbf{Algorithm 1}.

\begin{algorithm}  \caption{The training process of the BDQ based online optimization algorithm for microgrid.}
\small
\begin{algorithmic}[1]
    \State Initialize the neural networks of the BDQ agent; Initialize experience replay memory; Set the total number of episode $N_e$ and the training frequency $f_n$, and set the training step $n_{step} = 0$.
    \For {$n_e \le N_e$}
    	\State {Randomly select a day of renewable energy and load 
    	\Statex \quad \ \ sequences from the training data.}
		\For {$t = \Delta t, 2 \Delta t, \cdots, T$}
			\State {Get the current state information of the microgrid $s_t$, 
			\Statex \quad \quad \ \ \ and the previous $H$ hours of solar, wind, and load power.}
			\State {Compute the charge/discharge decisions of the BESSs
			\Statex \quad \quad \ \ \ using the BDQ agent. }
			\State {Recompute the charge/discharge decisions using $\epsilon$-greedy 
			\Statex \quad \quad \ \ \ policy.}
			\State {Check overcharge/overdischarge limits and get the opti- 
			\Statex \quad \quad \ \ \ mal decisions $P_d^{b,*}(t)$ ($d \in \{1, 2, \cdots, N\}$). }
			\State {Solve the OPF sub-problem to get the remaining de-
			\Statex \quad \quad \ \ \ cisions.}
			\State {Execute the optimal decisions $x_t$ to obtain the reward $r_t$,
			\Statex \quad \quad \ \ \ and calculate the next state of the system $s_{t+\Delta t}$.}
			\State Store the data $(s_t, x_t, r_t, s_{t+\Delta t})$ in the replay buffer.
			\If {($n_{step} \ \% \ f_n = 0 $)}
				\State Sample a minibatch of data from the replay buffer.
				\State {Update the main network weights of the BDQ agent 
				\Statex \quad \quad \quad \quad \ \ to minimize the loss function.}
				\State Update priorities of sampled data. 
				\State $n_{step} = n_{step} + 1$.
			\EndIf
			\State Update target network periodically.
		\EndFor 
		\If {($n_e \ \% \ 500 = 0 $)} \Comment {\textit{{\footnotesize Evaluate every 500 episodes.}}}
			\State Evaluate the optimization performance of the BDQ agent.
		\EndIf
		\State $n_e = n_e + 1$.
	\EndFor
	\State Return the well-trained BDQ agent parameters.
\end{algorithmic}
\end{algorithm}

\section{Case Study}
To demonstrate the effectiveness of the proposed BDQ based microgrid optimization algorithm, we tested the performance of the algorithm on a 6-bus microgrid system and a modified IEEE 33-bus microgrid system.
All the simulations are conducted on an Intel Core i7-8650U @1.90GHz Windows based computer with 16GB RAM.

\begin{figure}[h]\centering
\includegraphics[width=3.2in]{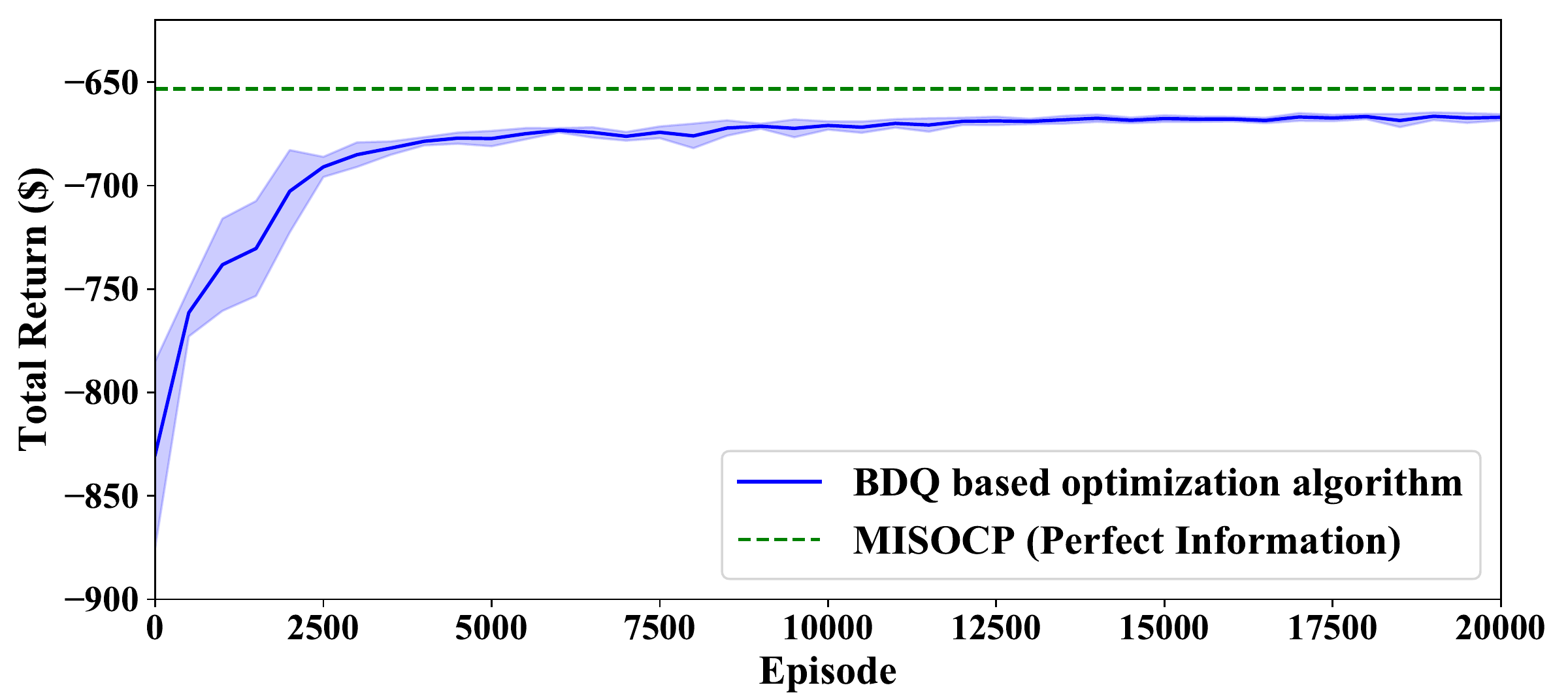}
\caption{The convergence process of the proposed BDQ based online optimization algorithm on the 6-bus microgrid system. Blue solid line indicates median returns across 5 separate training runs. The yaxis represents the average total returns for the 10 validation days.} \label{fig:case6}
\end{figure}

The topology of the adopted 6-bus microgrid can be found in \cite{Hang2020MCTS}, and the utilized training and testing dataset of solar power, wind power, load power, and electricity price are the same as in \cite{Hang2020MCTS}.
Although the 6-bus microgrid only contains one BESS, the proposed BDQ based optimization algorithm is also suitable.
The convergence process of the proposed algorithm is shown in Fig. \ref{fig:case6}.
From the result, the total returns optimized by the BDQ algorithm approaches the optimal value optimized by the MISOCP method under the condition of perfect information.
Note that the MISOCP method needs to know the accurate renewable generation and load power information of all the future time steps, so the optimal objective cannot be achieved in the online optimization process.

To test the online optimization performance of the proposed algorithm, we compared the BDQ based algorithm with several state-of-the-art online optimization algorithms.
Using the results optimized by myopic policy as the baseline, the performance improvement of the methods is shown in Table I.
We find that the proposed online optimization algorithm outperforms the Lyapunov optimization, ADP, and Deep Deterministic Policy Gradient (DDPG) based optimization algorithms.
Although the proposed algorithm performs worse than the MuZero based online optimization method proposed in \cite{Hang2020MCTS}, the MuZero based algorithm is difficult to apply in microgrids with multiple BESSs since the tree search space increases exponentially with the number of BESSs.

\begin{figure}[h]\centering
\includegraphics[width=3.5in]{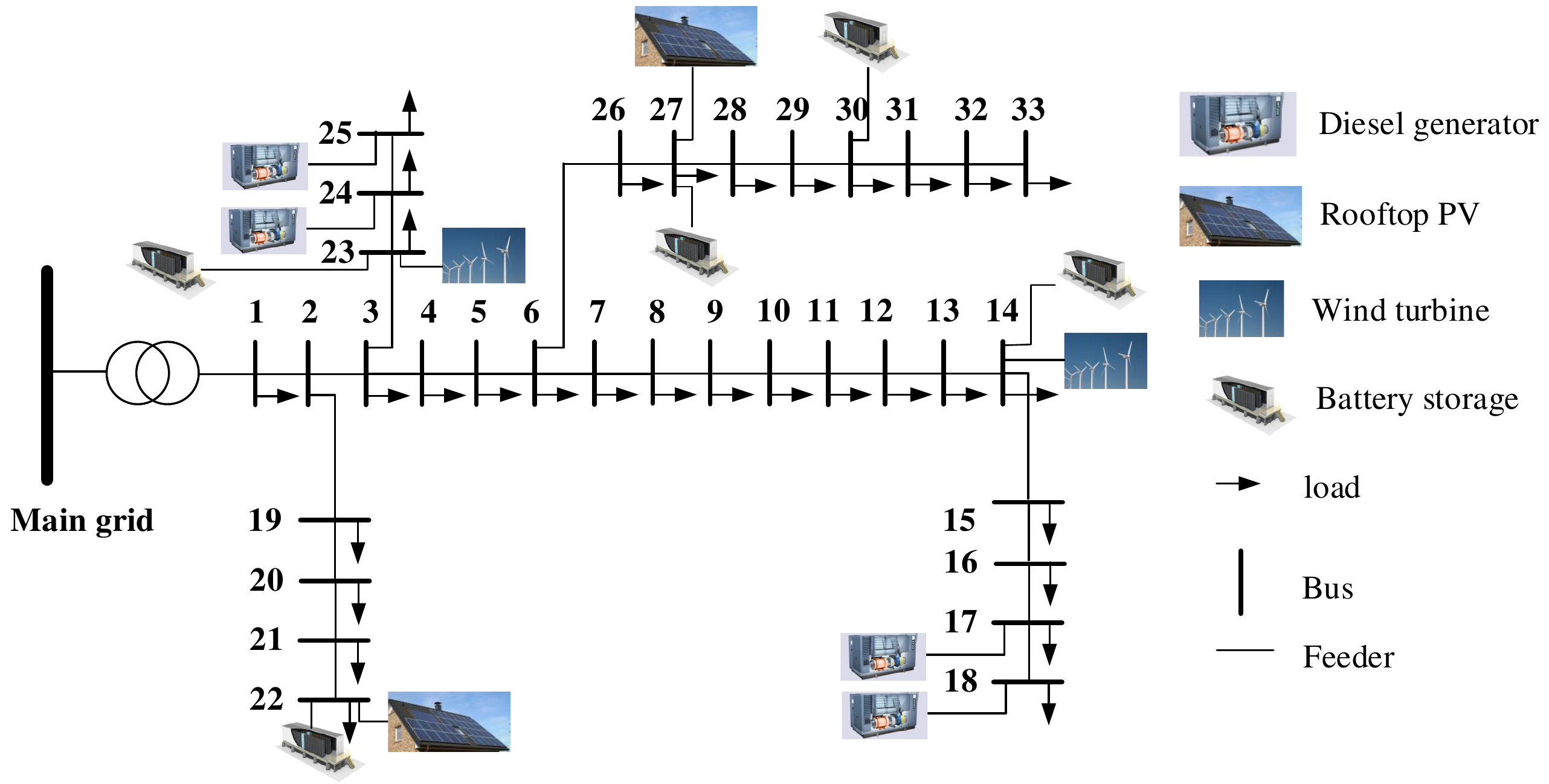}
\caption{The diagram of the modified IEEE 33-bus microgrid system.} \label{fig:case33}
\end{figure}

\begin{table}\centering
\begin{threeparttable}
\footnotesize
\newcommand{\tabincell}[2]{\begin{tabular}{@{}#1@{}}#2\end{tabular}}
\caption{The performance improvement of different methods compared to myopic policy on the 100-day testing dataset for the 6-bus microgrid system.}\label{Convergence1}
\setlength{\tabcolsep}{1mm}{
\begin{tabular}
{c|c|c|c|c|c} \hline \hline
\multicolumn{2}{c|}{\tabincell{c}{$Performance$ \\ $improvement$ }} &\tabincell{c}{$Mean$ } &\tabincell{c}{$Maximum$} &\tabincell{c}{$Minimum$} &\tabincell{c}{$Standard$ \\$deviation$} \\ \hline
\multirow{5}*{\centering \tabincell{c}{Online \\ methods}} &{\tabincell{c}{BDQ based \\ optimization}}  &8.52\%    &  19.94\%  &  3.19\%  &  2.65\%  \\
\cline{2-6}%
&{\tabincell{c}{MuZero based \\ optimization}} &9.30\%    & 16.68\%  &  5.28\% & 2.12\% \\ 
\cline{2-6}%
&{\tabincell{c}{Lyapunov \\ optimization}}  &3.76\%    &  9.89\%  &  1.93\%  &  1.65\%  \\ 
\cline{2-6}
&{\tabincell{c}{ADP}}  &6.57\%    &  14.78\%  &  4.16\%  &  1.92\%  \\ 
\cline{2-6}
&{\tabincell{c}{DDPG}}  &5.55\%    &  9.81\%  &  -8.41\%  &  4.16\%  \\ 
\cline{2-6} \hline
\multirow{1}*{}{\centering \tabincell{c}{Off-line \\ method}} &\tabincell{c}{MISOCP} &10.20\% &23.28\% &6.45\% &3.02\% \\
\hline
\hline
\end{tabular}}
\end{threeparttable}
\end{table}

\begin{table}\centering
\begin{threeparttable}
\footnotesize
\newcommand{\tabincell}[2]{\begin{tabular}{@{}#1@{}}#2\end{tabular}}
\caption{The performance improvement of different methods compared to myopic policy on the 100-day testing dataset for the modified IEEE 33-bus microgrid system.}\label{Convergence1}
\setlength{\tabcolsep}{1mm}{
\begin{tabular}
{c|c|c|c|c|c} \hline \hline
\multicolumn{2}{c|}{\tabincell{c}{$Performance$ \\ $improvement$ }} &\tabincell{c}{$Mean$ } &\tabincell{c}{$Maximum$} &\tabincell{c}{$Minimum$} &\tabincell{c}{$Standard$ \\$deviation$} \\ \hline
\multirow{5}*{\centering \tabincell{c}{Online \\ methods}} &{\tabincell{c}{BDQ based \\ optimization}}  &7.48\%    &  13.48\%  &  4.26\%  &  1.97\%  \\ 
\cline{2-6}%
&{\tabincell{c}{Lyapunov \\ optimization}}  &3.51\%    &  6.88\%  &  2.22\%  &  1.08\%  \\  
\cline{2-6}
&{\tabincell{c}{DDPG}}  &6.64\%    &  15.27\%  &  2.56\%  &  2.82\%  \\ 
\cline{2-6} \hline
\multirow{1}*{}{\centering \tabincell{c}{Off-line \\ method}} &\tabincell{c}{MISOCP} &10.20\% &23.40\% &6.58\% &2.98\% \\
\hline
\hline
\end{tabular}}
\end{threeparttable}
\end{table}

To validate the proposed algorithm's ability to solve the online scheduling of microgrids with multiple BESSs, the modified IEEE 33-bus microgrid system, which contains 5 distributed BESSs, was designed, as shown in Fig. \ref{fig:case33}.
Similarly, the online optimization performance of the proposed algorithm is evaluated and compared with the state-of-the-art methods.
The results are given in Table II.
Note that the MuZero based approach and look-up table ADP method face the curse of dimensionality due to the huge action space brought by multiple BESSs.
Thus, the comparable methods include only the Lyapunov optimization and the DDPG method.
It can be found that the proposed algorithm performs better than the compared methods, and the scheduling results of the proposed algorithm are near the optimal value.
\vspace{-0.2cm}
\section{Conclusion}
\vspace{-0.1cm}
A novel BDQ based online optimization algorithm for microgrids with multiple BESSs was proposed in this letter.
The proposed approach enables the linear growth of the total number of agent outputs with increasing BESSs, which provides great scalability and increases the applicability of the algorithm.
The simulations indicate that the online optimization performance of the proposed BDQ based approach outperforms the state-of-the-art online optimization methods, such as Lyapunov optimization, ADP, DDPG based method, and MuZero based method.
The easy implementation of the algorithm gives it a good application prospect.

\section*{Notice of Copyright}
This manuscript has been authored in part by UT-Battelle, LLC, under contract DE-AC05-00OR22725 with the US Department of Energy (DOE). The US government retains and the publisher, by accepting the article for publication, acknowledges that the US government retains a nonexclusive, paid-up, irrevocable, worldwide license to publish or reproduce the published form of this manuscript, or allow others to do so, for US government purposes. DOE will provide public access to these results of federally sponsored research in accordance with the DOE Public Access Plan (http://energy.gov/downloads/doe-public-access-plan).

\section*{Acknowledgments}

This material is based upon work supported in part by the US Department of Energy, Office of Energy Efficiency and Renewable Energy, Solar Energy Technologies Office Program,  under contract DE-AC05-00OR22725 and in part by CURENT, an Engineering Research Center funded by US National Science Foundation (NSF) and DOE under NSF award EEC-1041877.

\vspace{-0.3cm}
\bibliographystyle{IEEEtran}
\bibliography{IEEEabrv,BDQ_Microgrid_Ref}
\end{document}